\documentclass[aps,prb,twocolumn,preprintnumbers,amsmath,amssymb,superscriptaddress]{revtex4-1}
\usepackage{graphicx}
\usepackage{subfigure}
\usepackage{epsfig}
\usepackage{dcolumn}
\usepackage{bm}
\usepackage{ulem}
\usepackage{color}
\usepackage{multirow}
\usepackage{slashed}
\usepackage{hyperref}

\def\avg#1{\left\langle#1\right\rangle}

\def\be{\begin{equation}}       \def\ee{\end{equation}}
\def\bea{\begin{eqnarray}}      \def\eea{\end{eqnarray}}
\def\ba{\begin{array}}
\def\ea{\end{array}}
\def\bnum{\begin{enumerate} }
\def\enum{\end{enumerate}}

\def\nn{\nonumber}

\def\=>{\Rightarrow}
\def\>{\rightarrow}

\def\eye2{Fathbb{I}}

\renewcommand{\>}{\rangle}

\begin{document}

\title{Mass hierarchy in collective modes of pair-density-wave superconductors}

\author{Shao-Kai Jian}
\affiliation{Institute for Advanced Study, Tsinghua University, Beijing 100084, China}
\affiliation{Condensed Matter Theory Center, Department of Physics, University of Maryland, College Park, Maryland 20742, USA}

\author{Michael M. Scherer}
\affiliation{Institute for Theoretical Physics, University of Cologne, 50937 Cologne, Germany}

\author{Hong Yao}
\email{yaohong@tsinghua.edu.cn}
\affiliation{Institute for Advanced Study, Tsinghua University, Beijing 100084, China}
\affiliation{State Key Laboratory of Low Dimensional Quantum Physics, Tsinghua University, Beijing 100084, China}

\begin{abstract}
We study collective modes near the quantum critical point of a pair-density-wave (PDW) superconductor in two-dimensional Dirac systems. The fate of gaps of various collective modes is investigated by functional renormalization. For incommensurate PDW superconductors, we show that the gapless Leggett mode, protected by the emergent $U(1)$ symmetry, can induce an exponentially small Higgs mass compared to the superconducting gap. Further, for commensurate PDW superconductors, we find an emergent mass hierarchy in the collective modes, i.e. the masses of Leggett boson, Higgs boson, and the superconducting gap can differ by several magnitudes in the infrared. This may shed light to a mechanism underlying the hierarchy problem in the Standard Model of particle physics.
\end{abstract}

\date{\today}
\maketitle

\section{Introduction}

Collective modes in superconductors are among the most fascinating emergent phenomena in condensed matter physics~\cite{anderson1958a, anderson1958b}, and  are further related to the famous Anderson-Higgs mechanism~\cite{anderson1963, englert1964, higgs1964, kibble1964}.
In the case of charge-neutral particles, spontaneous breaking of  global~$U(1)$ symmetry provides massive amplitude and gapless phase fluctuations at low energies.
However, in the context of charged superconductors, i.e. electrons interacting with dynamic photons, the gapless Goldstone mode is ``eaten'' by gauge bosons resulting in massive transverse photons and the Meissner effect~\cite{schrieffer1964, martin1969, simons2010};
in this case, the amplitude mode is also known as Higgs mode.
Another collective mode -- the Leggett mode~\cite{leggett1966} -- appears in superconductors described by a superconducting (SC) order parameter with multiple components.
While the $U(1)$ transformation from charge conservation corresponds to a uniform phase shift in all components, the Leggett modes describe the relative phase fluctuations between different condensates.

One intriguing manifestation of a multicomponent superconductor is the pair-density-wave (PDW) superconductor whose order parameter transforms as a non-trivial representation of both $U(1)$ and lattice translation operations \cite{ferrell1964, larkin1964, berg2009, vishwanath2009, hui2010, moore2012, vafek2014, lee2014, fradkin2014, maciejko2014, jian2015, jian2017, kivelson2018, norman2018} (see Ref. \cite{PDW-ARCMP} for a recent review).
Recently, experimental evidences of PDW ordering in cuprate high-temperature superconductors were reported \cite{davis2018}.
The superconducting order parameter of a generic PDW reads
\bea
	\Delta(\vec r) = \Delta_+(\vec r) e^{i \vec Q \cdot \vec r} + \Delta_-(\vec r) e^{-i \vec Q \cdot \vec r},
\eea
where $\Delta_\pm$ correspond to two superconducting condensates that are related by time reversal (or inversion) symmetry.
Under a translation operation, the order parameters transform as
$\Delta_\pm \rightarrow e^{\pm iQ} \Delta_\pm$, where $Q= \vec Q \cdot \vec r$. 
Therefore, the phases of the two complex order parameters manifest themselves as order parameters of global charge conservation and translation symmetry.
It also manifests in secondary orders induced by the PDW, i.e., the $2\vec Q$ charge-density wave (CDW) $\rho_{\mathrm{CDW}}\!\sim\! \Delta_+ \Delta_-^\ast$ and the charge-$4e$ SC $\Delta_{4e} \!\sim\! \Delta_+ \Delta_-$~\cite{berg2009}.
Upon the transformation $\Delta_\pm \!\rightarrow\! e^{i \theta_\pm} \Delta_\pm$, $\rho_{\mathrm{CDW}}\!\rightarrow\! e^{i(\theta_+ - \theta_-)} \rho_{\mathrm{CDW}}$ and $\Delta_{4e} \!\rightarrow\! e^{i(\theta_+ + \theta_-)} \Delta_{4e}$.

It is clear that the induced CDW order is proportional to the Josephson coupling between the two condensates described by $\Delta_+$ and $\Delta_-$.
For incommensurate momentum $Q$, any local Josephson coupling, i.e., $(\Delta_+^\ast \Delta_-)^N$ for any integer $N$, is forbidden by the translational symmetry of the Landau theory and consequently, in addition to the global charge conservation, another $U(1)$ symmetry which is characterized by the phase difference between the two condensates emerges.
Consequently, the Leggett mode is gapless which is protected by the emergent $U(1)$ symmetry in the incommensurate PDW phase.
Surprisingly, we find that the fluctuations of the gapless Leggett mode can dramatically renormalize the mass of Higgs mode.
Specifically, we show that the Higgs mass is exponentially small compared to the superconducting gap -- i.e. the gap in fermion spectrum -- at low energies, and thus opens up the possibility of a detectable Higgs mode~\cite{klein1980, matsunaga2013, endres2012, sherman2015, behrle2018} in PDW superconductors.

On the order hand, for commensurate momentum $\vec Q$ with commensurability $N$, the minimum integer satisfying $2NQ= 2\pi\times$integer, the emergent $U(1)$ symmetry mentioned above is lowered to a discrete $Z_N$ symmetry and the Landau free energy can then allow the following Josephson coupling term at order $N$, $h[(\Delta_+ \Delta_-^{\ast})^N+ \mathrm{H.c.}]\!=\!2h|\Delta_+\Delta_-|^N \cos N(\theta_+-\theta_-)$,
where $h$ is a constant. In this situation, the Leggett mode will obtain a mass proportional to the strength of the order-$N$ Josephson coupling $J_N\propto 2h|\Delta_+\Delta_-|^N$.
As a result, the commensurability $N$ provides a knob to tune the mass of the Leggett mode:
for larger commensurability (larger $N$), the Josephson coupling is more irrelevant, and the Leggett boson mass gets smaller at low energies.
In 2+1 dimensions, the Josephson coupling is \textit{dangerously} irrelevant for $N \ge 3$, which will result in an interesting hierarchy~\cite{delamotte2018, yao2017, delamotte2015, scherer2018} of the various masses of collective modes as we will show below.

In the following, we implement a functional renormalization group (FRG) approach~\cite{wetterich1993, ellwanger1993, morris1994} to investigate collective modes in both incommensurate and commensurate PDW superconductors in Dirac systems.
The FRG is a nonperturbative approach to evaluate the effective action -- namely, the one-particle irreducible generating functional -- at any energy scale below the cutoff.
Importantly, it allows to study generic potential functions irrespective of whether they are perturbatively renormalizable or beyond~\cite{gies2013,borchardt2015,reichert2017}.
Thus, the FRG method is a suitable approach for the investigation of bosonic collective modes, where the effective potential is crucial for the determination of various gaps in the symmetry-broken phase.

\section{PDW state in honeycomb Dirac semimetals}
\begin{figure}[t]
	\subfigure[]{\label{fixedPoints}\includegraphics[height=2.6cm]{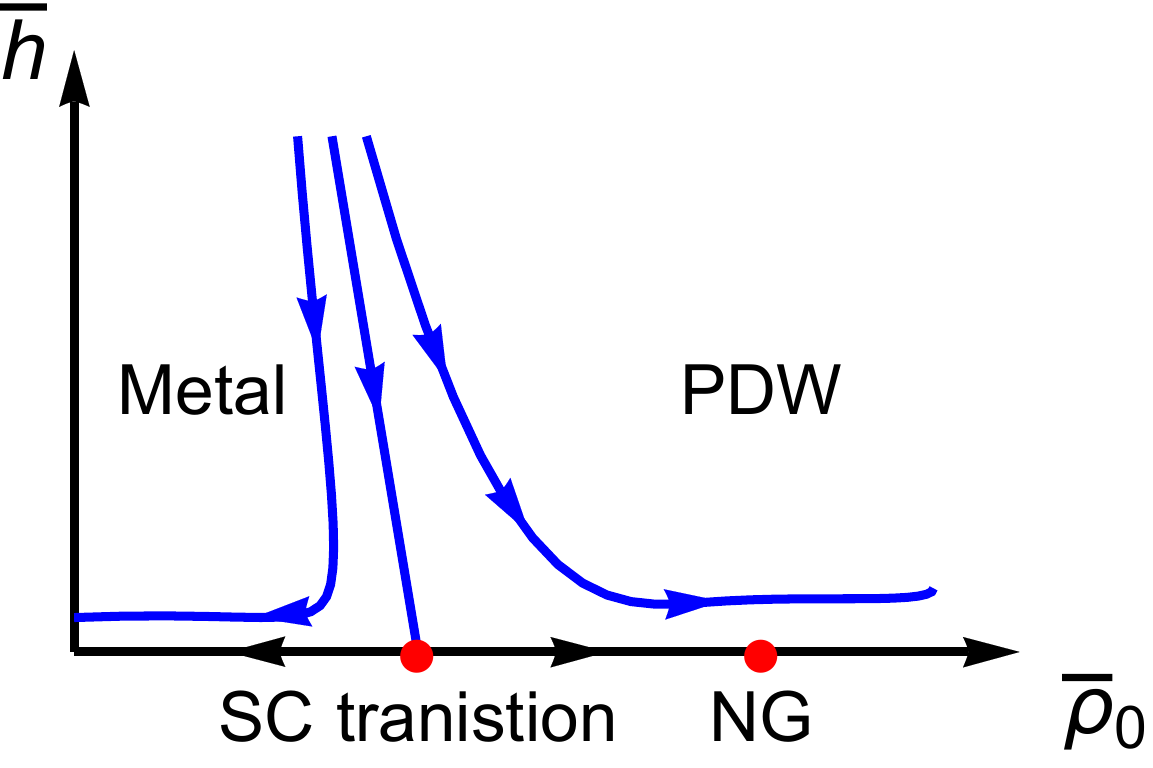}}
	\subfigure[]{\label{Nlambda2}\includegraphics[height=2.6cm]{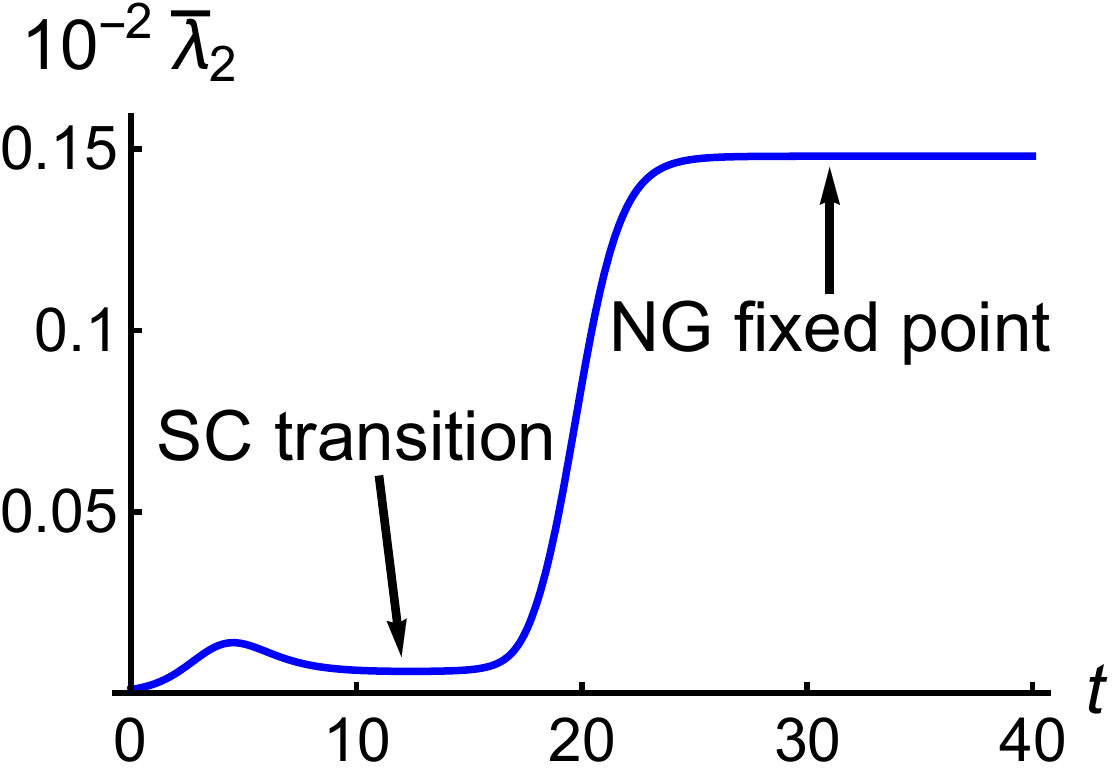}} \\
	\subfigure[]{\label{Ng2}\includegraphics[height=2.6cm]{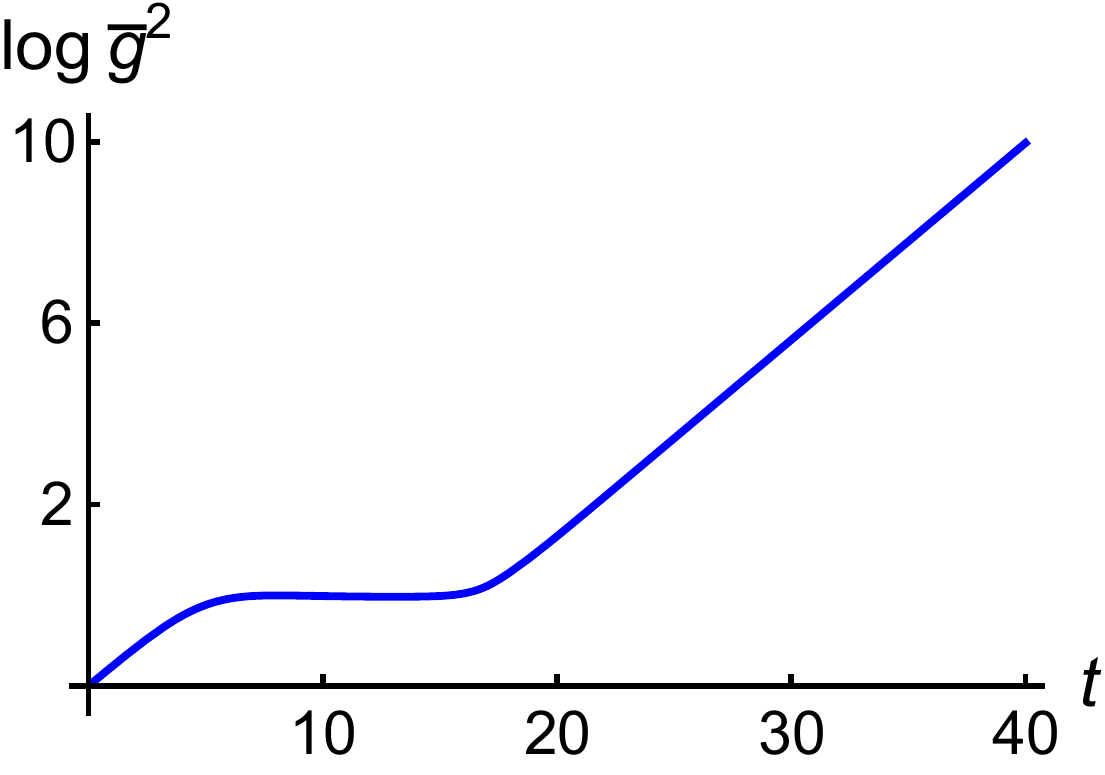}}
	\subfigure[]{\label{lambda11}\includegraphics[height=2.6cm]{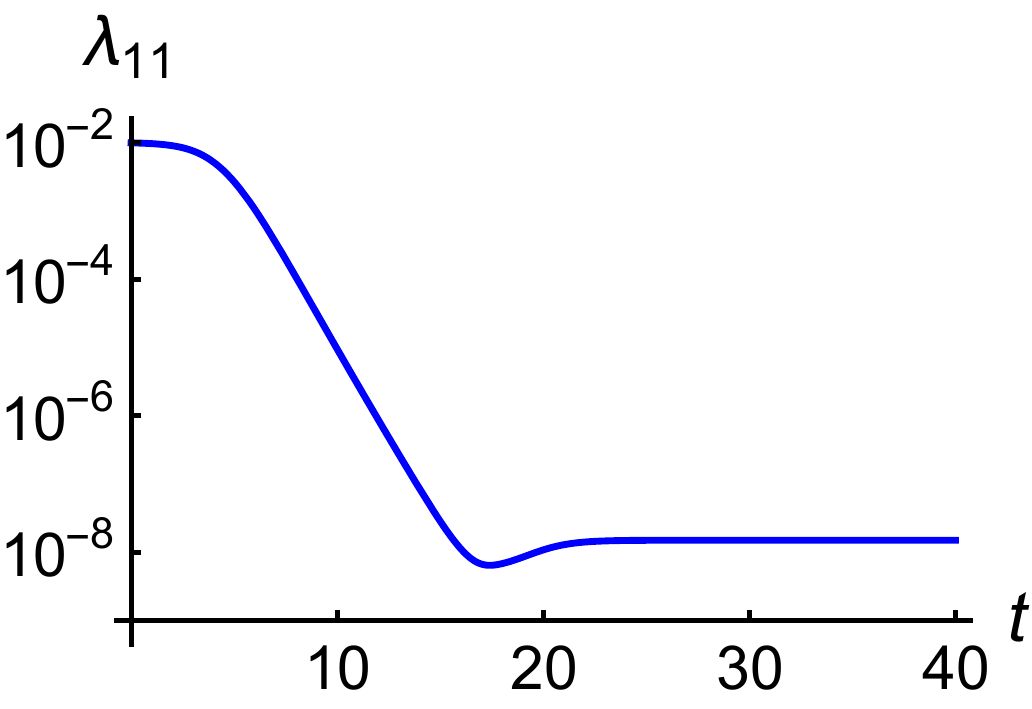}}
	\caption{\label{Nflow} (a) Schematic flow diagram near the phase transition between a (semi)metal and a PDW superconductor. The two axes represent the tuning parameter and the Josephson coupling strength, respectively. There are two fixed points, indicating by red points, corresponding to the superconducting (SC) transition point and the Nambu-Goldstone (NG) fixed point. The black and blue arrows indicate the flow for the incommensurate and commensurate PDW states, respectively, distinguished by whether the Josephson coupling is vanishing or not. (b), (c), and (d) are the flow diagrams of the potential coefficients in the incommensurate PDW phase.
The horizontal axis represents the flow parameter $t \equiv \log\Lambda_0/\Lambda$.
}
\end{figure}

%
We consider the PDW state of spinless fermions on a honeycomb lattice close to a quantum phase transition~\cite{jian2015, sachdevbook} as a primary example (see Fig.~\ref{fixedPoints} for a schematic phase diagram).
The half-filled honeycomb lattice hosts two Dirac cones at $K$ and $K'$ in the Brillouin zone, which are referred to as valley degrees of freedom and are denoted by $n=\pm$.
We consider a finite intravalley pairing, i.e. $\Delta_n\sim  \avg{\psi_n \sigma^y \psi_n}\neq 0$, which breaks the translation symmetry of the underlying lattice.
Under translation of the primitive lattice constant Dirac fermions and the order parameters transform as
$\psi_\pm \rightarrow e^{\pm i K}  \psi_\pm$ and
$\Delta_\pm  \rightarrow e^{\pm i 2 K} \Delta_\pm$, where $K=2\pi/3$ and we set the lattice constant to unity. Thus, the intravalley pairing state is a PDW superconducting state with commensurability $N=3$.
Such a state can, for example, be realized in the honeycomb model with nearest- and next-nearest-neighbor interactions~\cite{jian2015}.

In the charged PDW phase described above, its low-energy physics is described by the abelian Higgs model
\bea\label{abelianHiggs}
    &&S\!=\! \int_x \frac{-F^2}{4e^2} \!+\! \sum_{n=\pm} \Big[ |(\partial_\mu-i A_\mu) \Delta_n|^2 \!+\! \frac{r}2 |\Delta_n|^2 \!+\! \frac{u}{4!} |\Delta_n|^4 \Big]\nn \\
	&&~~~~~~~ + \frac{u'}{4} |\Delta_+|^2 |\Delta_-|^2 + h' (\Delta_+^3 \Delta^{\ast3}_-+ \mathrm{H.c.}) + \cdots,
\eea
where $\int_x \equiv \int d^3 x$, $A$ and $F$ are the vector potential and the field strength, respectively, and $e$ is the effective charge of Cooper pairs. The particle-hole symmetry of superconductors rules out a linear kinetic term~\cite{nambu1960, varma2015}.
Note that in Eq.~\eqref{abelianHiggs}, the gapped fermions are ignored for simplicity as their inclusion does not qualitatively change the discussion of the Higgs mechanism.
In terms of amplitude and phase modes, $\Delta_\pm= \varphi_\pm e^{i \theta_\pm}$, the kinetic energy is expressed as 	$|(\partial_\mu-i A_\mu) \Delta_\pm|^2= |\partial_\mu \varphi_\pm+ i \varphi_\pm (\partial_\mu \bar\theta - A_\mu \pm \partial_\mu \theta) |^2$,
where $\bar\theta= (\theta_++ \theta_-)/2$ and $\theta= (\theta_+-\theta_-)/2$ corresponding to Goldstone and Leggett modes, respectively.
In unitary gauge, $A_\mu \rightarrow A_\mu+\partial_\mu \bar\theta $, the Goldstone mode is eaten by the gauge field via the Higgs mechanism, while the gauge field obtains a mass.
The mass of the gauge boson is set by the amplitude of the SC order parameter, i.e., $|\Delta_\pm|$, which is comparable to the SC gap of the Dirac fermions.
Thus, as far as the physics below the SC gap is concerned, both the eaten-up Goldstone mode and the gauge boson can be neglected~\cite{schmalian2018}.
Note that it is reasonable to neglect the eaten-up Goldstone mode and the gauge boson in both incommensurate and commensurate PDW phases, although the discussion above is focused on the PDW phase with commensurability $N=3$.

\section{Collective modes in incommensurate PDW}

An incommensurate PDW can occur, e.g., through intra-valley pairing in a nematic Dirac semimetal that breaks the $C_3$ symmetry of the underlying honeycomb lattice.
Without $C_3$ symmetry, the Dirac point is still locally stable, but the momentum of the Dirac point is no longer locked at $K$ or $K'$.
A possible example is the twisted bilayer graphene, where an intermediate $C_3$ nematic semimetal phase is proposed~\cite{senthil2018}.
At a generic momentum, the PDW is incommensurate with the underlying lattice and an additional $U(1)$ symmetry emerges as discussed above.
The three boson degrees of freedom, i.e., two Higgs modes $\varphi_\pm$ and one Leggett mode $\theta$, can be changed to three real bosons, $\phi_\pm$ and $\phi$. i.e., $\Delta_\pm= \varphi_\pm e^{\pm i \theta} \to \Delta_\pm=\phi_\pm \pm i \phi$.

Now, we are ready to write down the bare action for the incommensurate PDW state with two Dirac fermions $S=S_0+ S_1$ with
\bea
	&&S_0= \int_x  \frac{(\partial \phi)^2}2  + \sum_{n=\pm}\bigg[\frac{(\partial \phi_n)^2}2  +   \Psi^\dag_n \mathcal{H}_n \Psi_n \bigg], \\
	&&S_1 = \int_x \lambda_{11}(\rho_+ -\rho_-)^2 + \sum_{n=\pm} \bigg[\frac{\lambda_2}2 (\rho_n - \rho_0 )^2 +  \nn \\
	&& ~~~~~+ \frac{\lambda_3}6 ( \rho_n- \rho_0 )^3 +  g \Psi^\dag_n  \sigma^y ( \phi_n \mu^x+ n \phi \mu^y) \Psi_n \bigg], \label{bareAction}
\eea
where $\Psi_\pm$ is the Dirac fermion in Nambu space with Pauli matrices $\mu^i$ and $\sigma^i$ acting on Nambu space and Dirac space, respectively. In Eq.~\ref{bareAction}, the $\rho$'s terms concern the bosonic action while the last term originates from Yukawa coupling in Nambu space. $\mathcal{H}_\pm=-i \omega \pm k_x \sigma^x+ k_y \sigma^y \mu^z$ is the kinetic term of the fermions and we have further introduced $\rho_\pm \equiv \frac12 (\phi^2_\pm + \phi^2)$. $\lambda_i$ characterizes the boson potential and $g$ is the Yukawa coupling. We consider a time-reversal invariant PDW phase, so the minimum of the potential is chosen to be located at $\phi_{\pm,\text{min}} = \sqrt{2\rho_0}$ and $\phi_\text{min}=0$.
In Eq.~\eqref{bareAction}, there is no Josephson coupling because of the incommensurability of the PDW phase under consideration and the emergent $U(1)$ symmetry renders the Leggett mode massless.

\section{Functional RG analysis}

We use the FRG approach to study the superconducting gap and the masses of the collective modes. The exact flow equation~\cite{wetterich1993} reads
$\partial_\Lambda \Gamma = \frac12 \text{Tr}[\partial_\Lambda R (\Gamma^{(2)} +R)^{-1} ]$,
where $\Gamma$ denotes the flowing effective action with energy scale $\Lambda$ and $\Gamma^{(2)}$ is the second functional derivative of the effective action with respect to boson and fermion fields. Furthermore, $R$ is a suitable cutoff function.
We implement the extended local potential approximation (LPA${}^\prime$) considering the following ansatz of effective action $\Gamma= \Gamma_0+\Gamma_1$ with
\begin{align}
	\Gamma_0 &= \int_x Z_b   \frac{(\partial \phi)^2}2  + \sum_{n=\pm}\bigg[Z_b \frac{(\partial \phi_n)^2}2  +  Z_f \Psi^\dag_n \mathcal{H}_n \Psi_n \bigg],\nn\\
	\Gamma_1 &= \int_x  \lambda_{11}(\rho_+ - \rho_-)^2 + \sum_{n=\pm} \bigg[ \frac{\lambda_2}2 (\rho_n - \rho_0 )^2 \nn\\
	 &\quad + \frac{\lambda_3}6 ( \rho_n- \rho_0 )^3 +  g \Psi^\dag_n  \sigma^y ( \phi_n \mu^x+ n\phi \mu^y) \Psi_n \bigg],
\end{align}
where $Z_i, i\in \{b,f\},$ are field renormalization factors.
We implicitly assume Higgs and Leggett modes have same field renormalization factor. Note that the fields $\phi_\pm$, $\Psi_\pm$, and $\phi$ in the effective action are expectation values and different from the fields in the bare action. For notational convenience we use same symbols.

With cutoff functions~\cite{litim2002} $ R_b =Z_b (\Lambda^2- k^2) \theta(\Lambda^2- k^2)$ and $R^f_n =Z_f  \mathcal{H}_n  r_f\Big(\frac\Lambda{k} \Big)$ where $r_f(x)= (x-1)\theta(x^2- 1)$, the flow equation for the bosonic potential $U$ reads
\bea
	\partial_\Lambda U &=&\frac{K_D Z_b \Lambda^{D+1}}{D}\bigg[ \sum_n \frac{ 1- \frac{\eta_b}{D+2}}{Z_b \Lambda^2+ m_n^2} +\frac{1- \frac{\eta_b}{D+2}}{Z_b \Lambda^2+ m_L^2} \bigg]  \nn \\
	&&~ - 2\frac{K_D Z_f^2\Lambda^{D+1}}{D}\sum_n \frac{1 - \frac{\eta_f}{D+1}}{Z_f^2 \Lambda^2 + 2 g^2 \rho_n}\,,  \label{potentialFlow}
\eea
where $D$ is the spacetime dimension and $K_D^{-1}= 2^{D-1} \pi^{D/2} \Gamma(D/2)$. The RG flow of $U$ can be projected to the flow of the minimum of the potential $\rho_0$ and the interaction coefficients $\lambda_{i}$. Their dimensionless versions are given as  $\bar \rho_0 \equiv Z_b \Lambda^{2-D} \rho_0$, $\bar\lambda_2 \equiv Z_b^{-2} \Lambda^{D-4} \lambda_2$, $\bar g^2 \equiv Z_f^{-2} Z_b^{-1} \Lambda^{D-4} g^2, \cdots$. The mass terms appearing in Eq.~\eqref{potentialFlow} and the following flow equations are evaluated at generic field configurations, $m_\pm= m_\pm(\phi_\pm, \phi)$ and $m_L= m_L(\phi_\pm, \phi)$.
At the minimum of the potential, two Higgs modes are given by $m_+^2=2\lambda_2 \rho_0 $ and $m_-^2=2\lambda_2 \rho_0 + 8 \lambda_{11} \rho_0$,
while the Leggett mode remains massless, $m_L=0$, due to the emergent $U(1)$ symmetry. The superconducting gap is $\Delta^2 =2g^2 \rho_0$.
The flow equation for the Yukawa coupling reads
\bea
	\partial_\Lambda g^2 &=& \frac{g^4K_D \Lambda^{D+1}}{2D(Z_f^2 \Lambda^2 + \Delta^2)}\bigg[ \frac{\big(1-\frac{\eta_f}{D+1}\big) Z_f^2}{Z_f^2 \Lambda^2 + \Delta^2} \Big(\frac1{Z_b^2 \Lambda^2 + m_+^2} \nn \\
	&+& \frac1{Z_b^2 \Lambda^2 + m_-^2}
	- \frac2{Z_b^2 \Lambda^2 + m_L^2}\Big)+\Big( \frac{Z_b\big(1 - \frac{\eta_b}{D+2}\big)}{(Z_b^2 \Lambda^2 + m_+^2)^2}  \nn \\
	&+& \frac{Z_b\big(1 - \frac{\eta_b}{D+2}\big)}{(Z_b^2 \Lambda^2 + m_-^2)^2} - \frac{2Z_b\big(1 - \frac{\eta_b}{D+2}\big)}{(Z_b^2 \Lambda^2 + m_L^2)^2}\Big) \bigg]\,.
\eea
Note that in the symmetry-breaking phase, the renormalization of Yukawa coupling $g^2$~\cite{metzner2008, metzner2010} is not vanishing. Finally, the anomalous dimensions are related to field renormalization factors by $\eta_i=- \Lambda Z_i^{-1} \partial_\Lambda Z_i $,
\bea
	\eta_b &=& \frac{8 K_D \lambda_2^2 \rho_0Z_b \Lambda^{D+2}}{D(Z_b\Lambda^2 + m_+^2)^2 (Z_b \Lambda^2 + m_L^2)^2} + \frac{8K_Dg^2 \Lambda^{D} Z_f^2}{Z_bD} \nn \\
	&\times& \Big[\frac{\big(D+2-\frac{4\eta_f}{D-1}\big) Z_f^2 \Lambda^2}{(D-2)(Z_f^2 \Lambda^2 +  \Delta^2)^3}- \frac{\frac34-\frac{\eta_f}2}{(Z_f^2 \Lambda^2 +  \Delta^2)^2 }\Big], \label{ZbFlow} \\
	\eta_f &=& \Big(1- \frac{\eta_b}{D+1} \Big) \frac{8 K_D Z_b \Lambda^{D+2} g^2}{D(Z_f^2 \Lambda^2+  \Delta^2)} \nn \\
	&\times& \Big[ \frac12  \sum_n \frac1{(Z_b \Lambda^2 + m_n^2)^2} + \frac1{(Z_b \Lambda^2 + m_L^2)^2} \Big] \label{ZfFlow}\,.
\eea

In Fig.~\ref{Nflow}, we show the flow diagrams for the couplings $\bar \lambda_2, \lambda_{11}$ and $\bar g^2$ as a function of flow parameter $t= \log\Lambda_0/\Lambda$. Here, $\Lambda_0$ is the cutoff energy of the bare action. The initial values of the RG flow are chosen in the PDW regime close to the transition point.
The flow diagram of $\bar \lambda_2 $, Fig.~\ref{Nlambda2}, shows two plateaus corresponding to the PDW transition point and the Nambu-Goldstone (NG) fixed point of the broken $U(1)$ symmetry owing to the incommensurability.
Note that the PDW transition point is a critical point, while the NG fixed point is a stable fixed point characterizing the gapless Leggett modes. The flow diagram of $\bar g^2$, Fig.~\ref{Ng2}, only shows the PDW transition plateau, because the fermions are gapped out in the SC phase and decouple from the low energy sector at the NG fixed point. Thus, the flow of $\bar{g}^2$ is set by its canonical dimension at the NG fixed point. The flow diagram of the dimensionful $\lambda_{11}$ shows its irrelevance at the critical point~\cite{jian2015}.

\begin{figure}[t]
	\subfigure[]{\label{Nmass}\includegraphics[height=2.6cm]{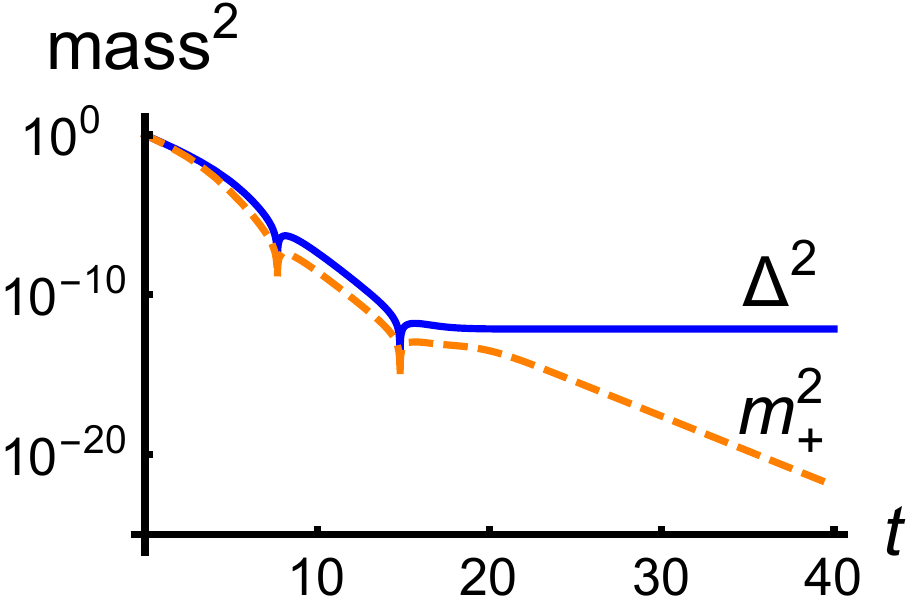}}~~~~
	\subfigure[]{\label{Nratio}\includegraphics[height=2.6cm]{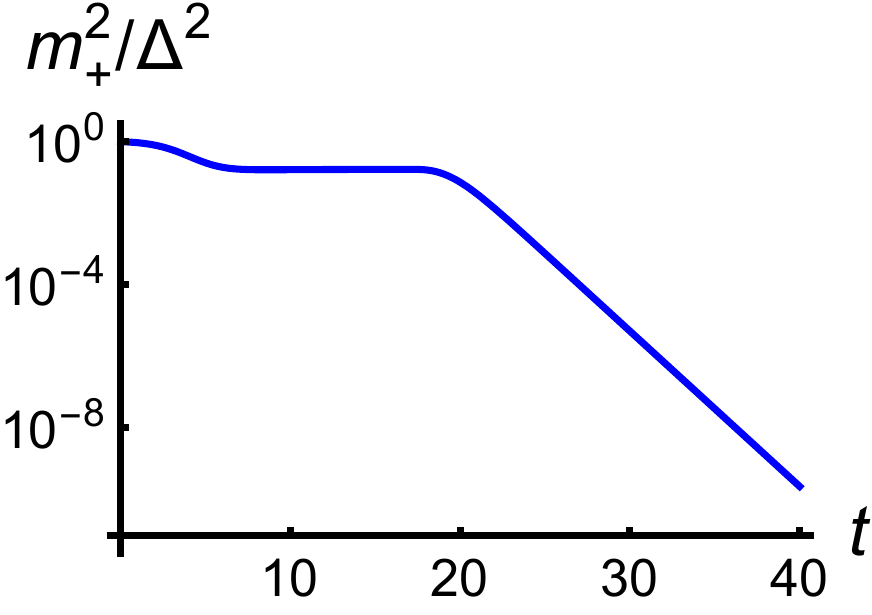}}
	\caption{The flow diagrams of the superconducting gap and mass of amplitude mode at incommensurate PDW phase. We have scaled the initial value to 1. (a) shows the flow of $\Delta^2$ and $m_+^2$. Notice that the flow quantities have dimension mass squared. (c) shows the flow of the ratio $m_+^2/\Delta^2$.}
\end{figure}

After identifying two fixed points, we can study in more details the flow of the SC gap and Higgs boson mass towards low energies and we focus on one of the Higgs modes, i.e., $m_+^2$, for simplicity.
We show the flow of the dimensionful squared masses $\Delta^2$ and $m_+^2$ in Fig.~\ref{Nmass}.
At the energy scale controlled by the PDW transition point, the RG flows of SC gap and Higgs boson mass are almost identical.
At lower energies, the physics is controlled by the NG fixed point: while $\Delta^2$ stops flowing because it decouples from the low energy sector, $m_+^2$ continues to flow due to fluctuations of the massless Leggett mode. Eventually, $m_+^2$ flows to zero at extremely low-energies~\cite{supp}.
Fig.~\ref{Nratio} shows the flow of the ratio between SC gap and Higgs boson mass. After the system enters the energy scale controlled by the NG fixed point, the Higgs mass gets exponentially smaller compared to the SC gap at low energy. This provides a robust energy window where the Higgs modes are detectable in an incommensurate PDW superconductor. The response of Higgs mode to external probe is similar to that in neutral SC/superfluid~\cite{supp, arovas2011}.

\section{Collective modes in commensurate PDW}

We now study the case of commensurate PDW.
Due to the commensurability $N=3$, we add a Josephson coupling term $S_J\propto \Delta_+^3 \Delta_-^{\ast3}+ \mathrm{H.c.}$ to the action which couples the two SC condensates.
In terms of real bosons $S_J$ reads
\bea\label{Z6}
	S_J = h\int_x \Big( 8\sum_n \rho_n^3 - [(\phi_++i\phi)^3 (\phi_-+i\phi)^3+ \mathrm{H.c.}] \nn\Big)\,. \nn
\eea
We added $\sum_n \rho_n^3$ such that $S_J$ is nonnegative and the minimum of the potential is still at $\phi_{\pm,\text{min}}= \sqrt{2\rho_0}$, $\phi_\text{min}=0$.
Including a term corresponding to $S_J$ in the truncation, the flow equations of the potential and fermion anomalous dimension are same as Eqs. (\ref{potentialFlow}) and (\ref{ZfFlow}), except the masses are different due to the presence of the Josephson coupling.
The flow equation of boson anomalous dimensions is
\bea
	\eta_b &=& \frac{8 K_D Z_b \Lambda^{D+2}(\lambda_2+288 h \rho_0)^2 \rho_0}{D(Z_b\Lambda^2 + m_+^2)^2 (Z_b \Lambda^2 + m_L^2)^2}   + \frac{8K_Dg^2 \Lambda^{D} Z_f^2}{DZ_b}\nn\\
	&&\times
	\Big[\frac{\big(D+2-\frac{4\eta_f}{D-1}\big) Z_f^2 \Lambda^2}{(D-2)(Z_f^2 \Lambda^2 + \Delta^2)^3} - \frac{\frac34-\frac{\eta_f}2}{(Z_f^2 \Lambda^2 + \Delta^2)^2 } \Big]. \label{ZbFlow2}
\eea
%
The masses of Higgs mode and Leggett mode are given by $m_+^2=2\lambda_2 \rho_0$ and
$m_L^2=288 h \rho_0^2$, respectively. Note that the Leggett boson mass is proportional to strength of the Josephson coupling.

\begin{figure}[t]
	\subfigure[]{\label{lambda2}\includegraphics[height=2.8cm]{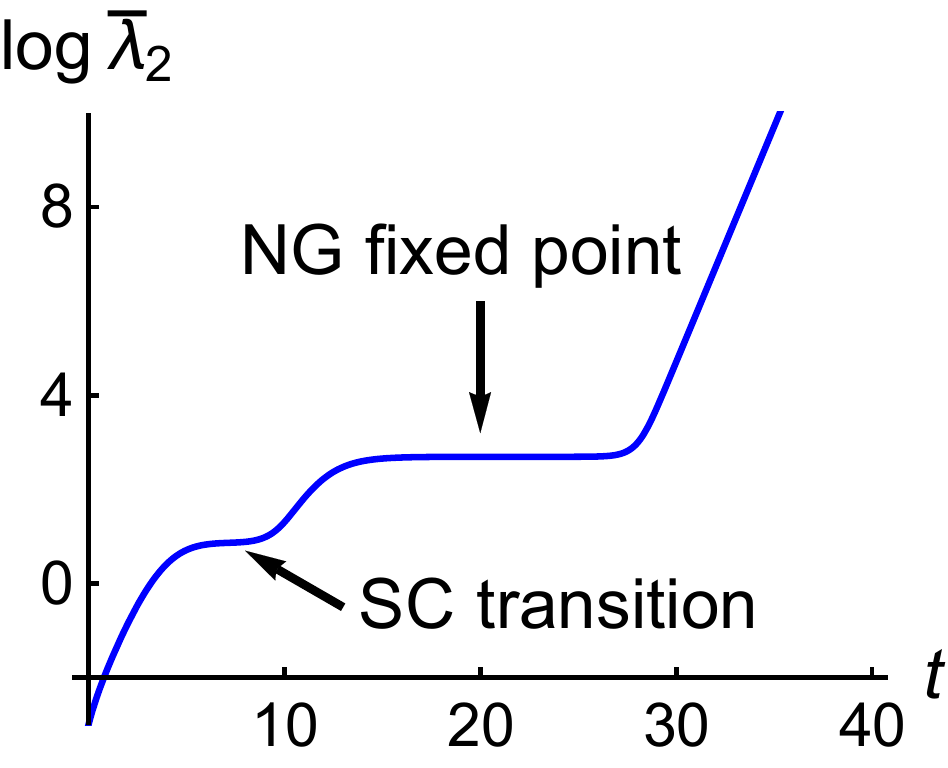}}
	\subfigure[]{\label{mass}\includegraphics[height=2.8cm]{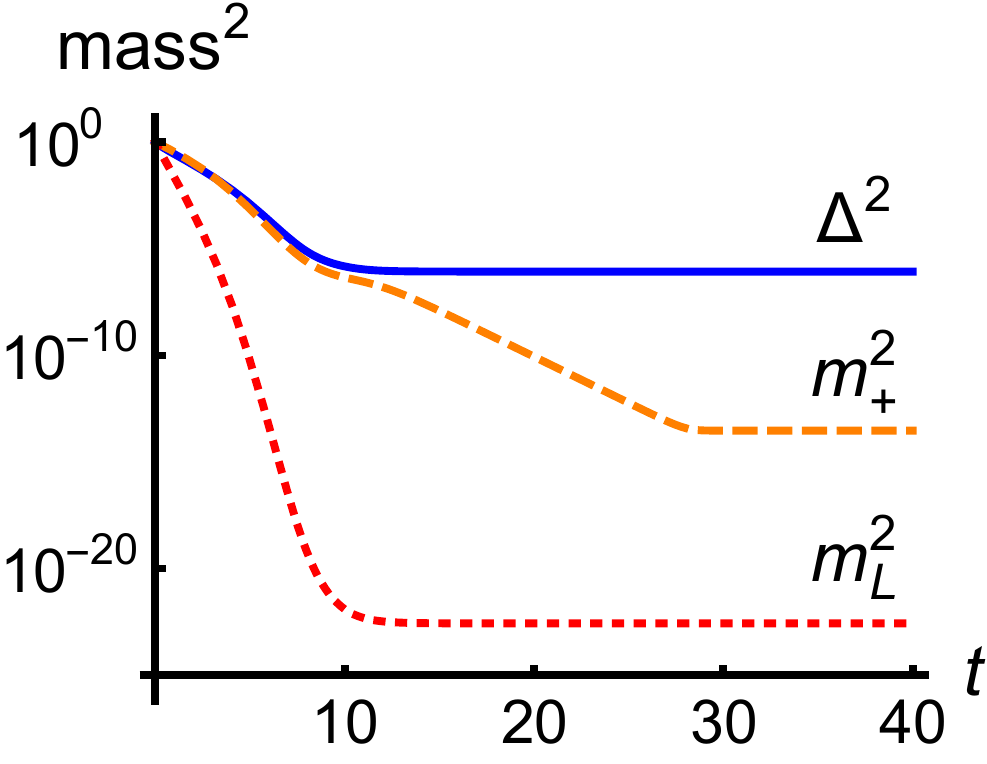}}
	\caption{\label{flow} The flow diagrams of the potential coefficient $\bar \lambda_2$, SC gap, and masses of various collective modes at commensurate PDW phase. We have scaled the initial value to 1. (a) shows the flow of dimensionless coefficient $\bar \lambda_2$ $\Delta^2$. (b) shows the flow of superconducting gap, Higgs boson mass, and Leggett boson mass denoted by $\Delta^2$, $m_+^2$, and $m_L^2$ respectively. Notice that the flow quantities have dimension mass squared. }
\end{figure}

To study the flow of collective modes in the commensurate PDW, we set initial values in the PDW regime close to the transition point. Fig.~\ref{lambda2}, the flow diagram of $\bar \lambda_2$, shows two fixed points corresponding to the PDW transition point and the NG fixed point similar to that in incommensurate PDW state.
However, the NG fixed point is unstable in the commensurate PDW state due to the run-away flow of $\bar \lambda_2$ after the NG fixed point.
This behavior originates in the Josephson coupling which is dangerously irrelevant at the PDW transition point and triggers the run-away flow of $\bar \lambda_2$.
In Fig.~\ref{mass}, we show the flow diagrams of the SC gap and the masses of the collective modes. In the energy range controlled by PDW transition point, the flows of SC gap $\Delta^2$ and Higgs mass $m_+^2$ are identical.
More interestingly, the flow of the Leggett boson mass $m_L^2$ is faster, because of the irrelevance of $h$ at the PDW transition. Note that it is a robust feature insensitive to the initial values~\cite{supp}. Thus, it provides an energy window to detect the Leggett boson, i.e. the (dangerous) irrelevance of the Josephson coupling makes the Leggett mode detectable in commensurate PDW.

In the energy range controlled by the NG fixed point only Higgs boson mass continues to flow, similar to incommensurate PDW state, while both SC gap and Leggett boson mass stop running.
Finally, when the system reaches lower energies, all masses stop flowing and remain finite: unlike the incommensurate PDW state where the enhanced $U(1)$ symmetry protects gapless Leggett modes, there is no protected gapless mode in the commensurate PDW phase. The presence of two fixed points gives rise to an interesting emergent hierarchy of boson masses~\cite{yao2017}, which may shed light to a mechanism underlying the hierarchy problem in the Standard Model of particle physics~\cite{delamotte2018}.

\section{Concluding remarks}

By using the FRG method, we show that, (a) in an incommensurate PDW superconductor, the Higgs mass is exponentially smaller than the superconducting gap near the superconductor transition point, due to the gapless fluctuations of Leggett modes, and (b) in the commensurate PDW phase, the Leggett boson mass is finite but exponentially small compared to Higgs boson mass and superconducting gap, i.e., a mass hierarchy of the collective modes emerges.

While we studied the PDW state in Dirac semimetals as an explicit example, we note that the results are robust in nodeless PDW superconductors: in the incommensurate PDW state, the gapless Leggett modes --which are protected by the emergent $U(1)$ symmetry -- can strongly renormalize the Higgs mass. These findings are in general correct in nodeless incommensurate PDW superconductors. On the other hand, in commensurate PDW superconductors, the mass hierarchy between Higgs boson and Leggett boson masses relies on the fact that the Josephson coupling is dangerously irrelevant at the PDW transition point. This happens generically in PDW states with high commensurability. 

{\it Acknowledgement. ---} We thank Shuai Yin for helpful discussions. This work was supported in part by the NSFC under Grant No. 11825404 (S.-K.J. and H.Y.), the Simons Foundation via the It From Qubit Collaboration (S.-K.J.), the MOSTC under grant Nos. 2016YFA0301001 and 2018YFA0305604 (H.Y.), the Strategic Priority Research
Program of Chinese Academy of Sciences under Grant No. XDB28000000 (H.Y.), and Beijing Municipal Science \& Technology Commission under grant No. Z181100004218001 (H.Y.), and Beijing Natural Science Foundation under grant No. Z180010 (H.Y.). M.M.S was supported by the DFG, Projektnummer 277146847 - SFB 1238 (project C02, C04).

\begin{widetext}
\begin{center}
\bf APPENDIX
\end{center}


\subsection{Mechanism of the emergent mass hierarchy}

Here we would like to articulate more about the origin of the mass hierarchy between Leggett mode and Higgs mode in the commensurate pair-density-wave~(PDW) superconductor. In short, it is due to the irrelevance of Josephson coupling at criticality. More specifically, this mass hierarchy can be quantified by the inverse ratio of the masses of the Higgs mode $m_+^2\propto \lambda_2\rho_0$ and the Leggett mode $m_L^2\propto h_N \rho_0^{N-1}$, where $h_N$ is a Josephson coupling with commensurability $N$,
\begin{align}
	\frac{m_L^2}{m_+^2} \propto \frac{\bar h_N \bar\rho_0^{N-2}}{\bar \lambda_2}\,.
\end{align}
Near the critical point, the RG flows of the dimensionless $ \bar\rho_0$ and $\bar \lambda_2$ show plateaus, i.e. their values stay nearly constant. On the other hand, at the critical point the dimensionless $\bar h_N$ will flow to zero asymptotically, because, here, the Josephson coupling is irrelevant for large enough $N$ ($N\ge 3$). The longer the RG flow is dominated by the critical point, the smaller becomes the ratio $m_L^2/m_+^2$, independent of the initial value of the Josephson coupling. This can be controlled by tuning closer to criticality. Eventually, in the deep infrared, both masses freeze out and give a constant tiny but finite mass ratio --  the mass hierarchy.

The canonical scaling dimension of $\bar h_N$ is $(2-D)N+D$. For $D=3$ and $N=3$, $\bar h_N$ is marginal at tree level and quantum fluctuations have to be considered. Employing the non-perturbative functional RG approach, we find that the quantum fluctuations render the Josephson coupling irrelevant at the critical point, which is also in agreement with simpler perturbative arguments. Here, we add another non-perturbative reasoning, which further supports our finding for the PDW transition in the Dirac semimetal as considered in the manuscript:
The PDW transition for this specific system features emergent supersymmetry (SUSY)~\cite{jian2015}. More explicitly, it is described by two decoupled copies of $\mathcal{N}=2$ Wess-Zumino supersymmetric theory leading to an exact scaling dimension of complex boson, $[\Delta_\pm]=2/3$, and constraining the quantum fluctuation such that $[\Delta_\pm^N]=N\times [\Delta_\pm]$~\cite{strassler2003}. Since two valleys are decoupled at criticality, we have $[(\Delta_+^\ast \Delta_-)^N]=[(\Delta_+^\ast)^N]+[ (\Delta_-)^N]=4N/3$, and $[\bar h_N]=3-4N/3$. As a result, the mass ratio is
\bea
	\frac{m_L^2}{m_+^2} \propto \frac{\bar h_N(\Lambda_0) \bar\rho_0^{N-2}(\Lambda)}{\bar \lambda_2(\Lambda)} \big( \frac{\Lambda}{\Lambda_0} \big)^{3-4N/3},
\eea
where $\Lambda_0$ is the energy cutoff and $\Lambda$ represents the energy scale. Because $\bar\rho_0(\Lambda)$ and $\bar \lambda_2(\Lambda)$ stay roughly constant near the transition point, we can see the reason of mass hierarchy is due to the irrelevance of Josephson coupling. Moreover, the mass hierarchy is more apparent for the larger $N$.

We clearly exhibit this result by showing the renoramlization group flow of the mass for different choices of tuning parameters away from criticality, see Fig.~\ref{fig1}. This confirms that an initial value which is closer to the critical point induces a smaller the Leggett mass and mass ratio $m_L^2/m_+^2$ as argued above. Note the small increase of the mass ratio in the figure is due to the Nambu-Goldstone (NG) fixed point, which further decrease the Higgs mass but not Leggett mass.
\begin{figure}
	\subfigure[]{
		\includegraphics[width=5cm]{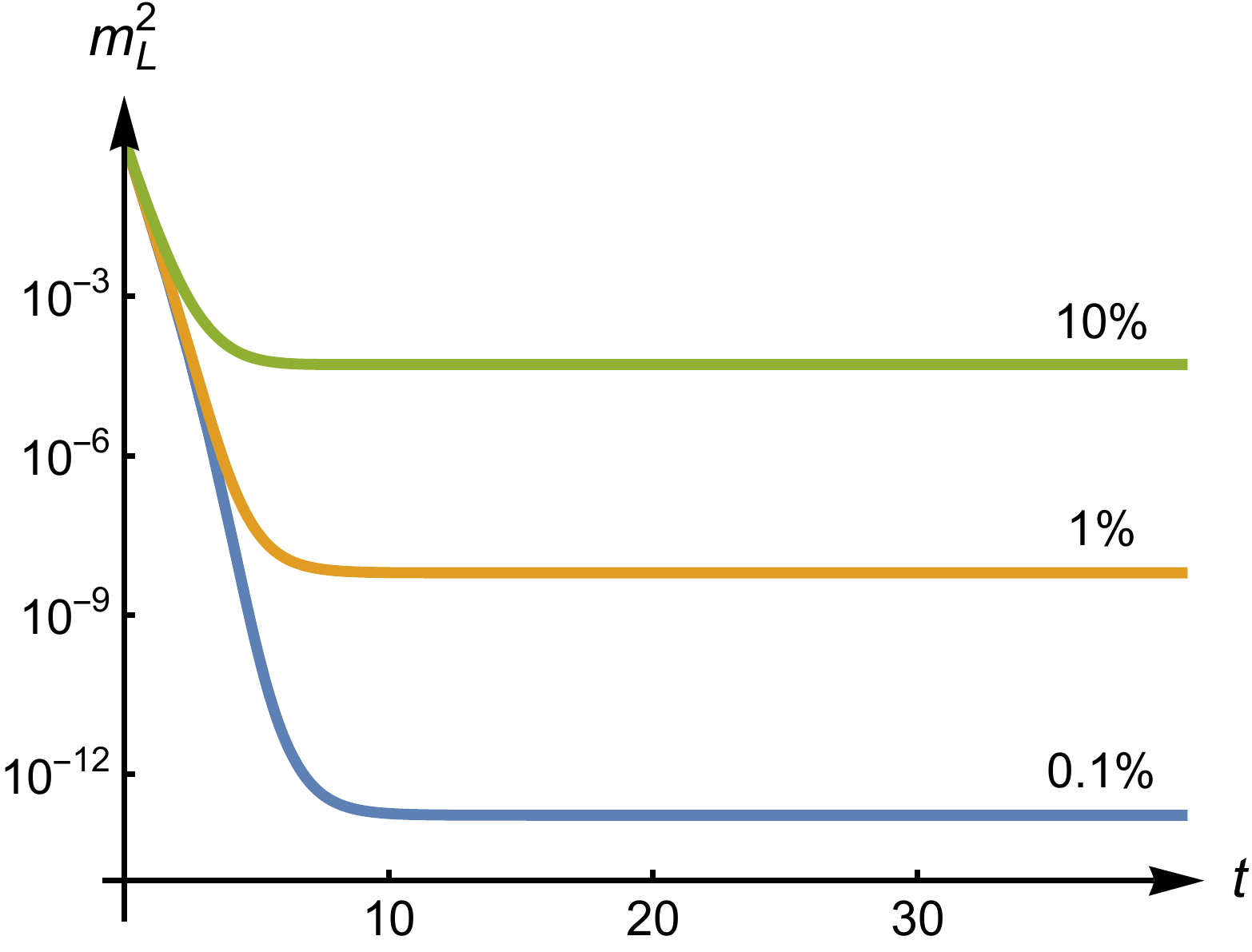}} \quad
	\subfigure[]{
		\includegraphics[width=4.8cm]{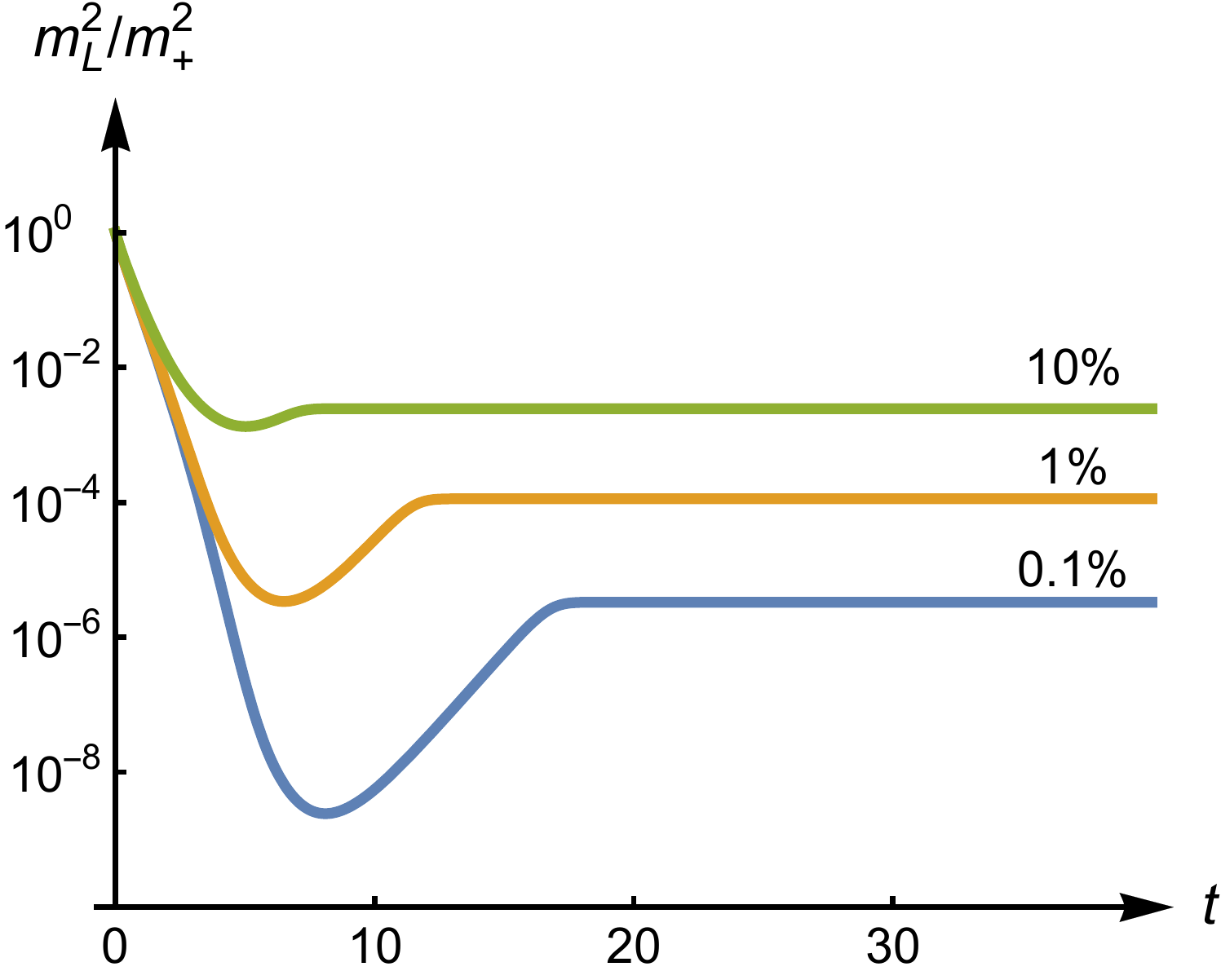}}
\caption{\label{fig1} The flow of the mass of the Leggett mode and the mass ratio for different choices of the initial parameters, namely, at $(\bar \rho_0-\bar \rho^\ast)/\bar \rho^\ast=$ 10\%, 1\% and 0.1\% away from critical point in the symmetry-broken phase. }
\end{figure}

The above results explain the mass hierarchy between the Leggett mode and the Higgs mode. Next, we provide a physical reason for the mass hierarchy of the Higgs mode and the superconducting (SC) gap. From Fig. 2 and Fig. 3 in main text, we can see that the mass hierarchy emerges near the NG fixed point. Because the mass ratio is $m_+^2/\Delta^2 \propto \bar\lambda_2/\bar g^2$, the hierarchy is due to the different IR behavior of $\bar\lambda_2$ and $\bar g^2$. The FRG equations near the incommensurate PDW transition are given by
\bea
	\Lambda \partial_\Lambda \bar \lambda_2 &=& - \bar\lambda_2 + \frac{3\bar \lambda_2^2}{\pi^2(1+2\bar\lambda\bar \rho_0)^3} - \frac{\bar g^2}{(1+2\bar g^2 \rho_0)^3} + \frac{2\bar\lambda_2^2}{3\pi^2}, \\
	\Lambda \partial_\Lambda \bar g^2 &=& - \bar g^2 + \frac{\bar g^4}{6\pi^2(1+2 \bar g^2 \bar \rho)^2} \Big[ \frac1{1+2\bar\lambda_2 \bar \rho_0}-1 +(1+2 \bar g^2 \bar \rho)\Big(\frac1{(1+2\bar\lambda_2 \bar \rho_0)^2}-1 \Big) \Big].
\eea
The above equation allows us to directly analyze the behavior of the system at the NG fixed point: Since the NG fixed point describes the symmetry broken phase, the relevant tuning parameter diverges, i.e. $\bar \rho_0 \rightarrow \infty$, and the RG equations reduce to
\bea
	\Lambda \partial_\Lambda \bar \lambda_2 &=& - \bar\lambda_2 + \frac{2\bar\lambda_2^2}{3\pi^2}, \\
	\Lambda \partial_\Lambda \bar g^2 &=& - \bar g^2.
\eea
The flow of $\bar g^2$ is set by its canonical dimension because the quantum corrections vanish due to the decoupling of the fermion from the low-energy sector. With the solutions of the RG equation at the NG fixed point,
\bea
	\bar\lambda_2(\Lambda) &=& \frac1{\frac2{3\pi^2}+(\frac1{\bar\lambda_2(\Lambda_0)}- \frac2{3\pi^2}) \frac{\Lambda}{\Lambda_0}}, \quad \bar g^2(\Lambda) = \bar g^2(\Lambda) \frac{\Lambda_0}\Lambda,
\eea
we can get the mass ratio
\begin{align}
\frac{m_+^2}{\Delta^2} \propto \frac{3\pi^2}{2 \bar g^2(\Lambda_0)} \frac{\Lambda}{\Lambda_0}\,.
\end{align}
Physically, the mass hierarchy is due to the fact that fermion decouples from the low-energy sector at the NG fixed point, while the Higgs mode does not.

\subsection{B. Observability of Higgs mode in incommensurate PDW superconductors}

In this section, we discuss the observability of Higgs modes in PDW superconductors (SC) in 2+1D. One necessary condition of a well-defined Higgs mode is Lorentz symmetry. In condensed matter physics, which typically provides a non-relativistic environment, the particle-hole symmetry in superconductors plays an essential role similar to Lorentz symmetry~\cite{varma2015}. In the incommensurate PDW SC, it is not obvious whether Higgs mode can be probed in experiments due to the gapless fluctuations of Leggett mode. The effective theory, i.e. Eqs. (3) and (4), in terms of Higgs and Leggett modes is
\bea
	\mathcal L &=&  \sum_n \Big[ (\partial \phi_n)^2 + (\lambda_2 \rho_0 + 2\lambda_{11} \rho_0) \phi_n^2 + \lambda_2 \frac{\rho_0}2 (\phi_n^3 + \phi_n \phi^2)\Big] + (\partial \phi)^2 + \frac{\lambda_2}8 \phi^4 \\
	&& -4\lambda_{11} \rho_0 \phi_+ \phi_- + \frac{ \lambda_{11}}4(\phi_+ - \phi_-)^2 (2\sqrt{2\rho_0} + \phi_+ + \phi_-).
\eea
Note that $\lambda_{11}$ only couples the Higgs modes. To address the effect of the gapless fluctuation of Leggett mode, one can set $\lambda_{11}=0$ for simplicity, i.e.
\bea
	\mathcal L &=&  \sum_n \Big[ (\partial \phi_n)^2 + (\lambda_2 \rho_0 + 2\lambda_{11} \rho_0) \phi_n^2 + \lambda_2 \frac{\rho_0}2 (\phi_n^3 + \phi_n \phi^2)  \Big] + (\partial \phi)^2 + \frac{\lambda_2}8 \phi^4.
\eea
The polarization operator of Higgs mode is given by
\bea	
	\Pi(q) = \frac14 \int \frac{d^Dk}{(2\pi)^D} \frac1{k^2(k+q)^2} = \frac14 \frac{1}{(4\pi)^{D/2}} \frac{2^{3-D} \sqrt{\pi} \Gamma(D/2-1)}{\Gamma{(D/2-1/2)}} \frac{\Gamma{(2-D/2)}}{q^{4-D}}\,.
\eea
Since $\Pi(q)=\frac1{32} \frac1{\sqrt{q^2}}$ in 2+1D, the self-energy suffers from an IR singularity, making the direct observation of longitudinal fluctuations difficult. Indeed, the spectral function of the longitudinal susceptibility at one-loop order is given by
\bea
	\chi''_{\phi_n \phi_n} = \frac{\pi}{4\sqrt{q^2 + \lambda_2 \rho_0}} \delta(\omega - \sqrt{q^2 + \lambda_2 \rho_0}) + \frac{\lambda_0^2 \rho_0}{128}\frac{1}{(\omega^2 - q^2 - \lambda_2 \rho_0)^2} \frac1{\sqrt{\omega^2 -q^2}} \Theta(\omega^2-q^2),
\eea
where $\Theta$ is the step function. The first term is the quasiparticle peak of the Higgs mode, and the second term comes form the decay to Leggett mode.
In the static limit, $\chi''_{\phi_n \phi_n} \sim \omega^{-1}$, making the Higgs peak difficult to be observed. Note that such a situation is ubiquitous for the amplitude mode in spontaneous continuous symmetry breaking in 2+1D; the gapless Goldstone mode fluctuations inhibit probing the longitudinal susceptibility~\cite{patashinskii1973, arovas2011}. In order to probe the signature of Higgs modes, one can instead consider the scalar susceptibility. The scalar is defined as $\delta\rho_n \equiv \frac1{\sqrt{2\rho_0}}(\frac12 |\Delta_n|^2 - \rho_0) =  \phi_n + \frac{\phi_n^2 + \phi^2}{2\sqrt{2\rho_0}}$, and the scalar susceptibility is
\bea
	\chi_{\delta \rho_n \delta \rho_n} =  \chi_{\phi_n \phi_n} + \frac1{\sqrt{2\rho_0}} (\chi_{\phi_n \phi^2} +\chi_{\phi_n \phi_n^2}) + \frac1{8\rho_0} (\chi_{\phi^2 \phi^2} + \chi_{\phi_n^2 \phi_n^2}) + \frac1{4\rho_0} \chi_{\phi_n^2 \phi^2}.
\eea
We calculate the susceptibility by a weak coupling expansion (loop expansion), and present the one-loop result. The singular loop diagrams shown in Fig. \ref{fig2} lead to the singular part of the susceptibility,
\bea
	\chi_{\delta \rho_n \delta \rho_n}^\text{singular}(q) = \lambda_2^2 \rho_0 \chi_0(q) \Pi(q) \chi_0(q) -\lambda_2 \chi_0(q) \Pi(q) + \frac1{4\rho_0} \Pi(q) = \frac{q^4}{4\rho_0(q^2+ \lambda_2 \rho_0)^2} \Pi(q),
\eea
where $\chi_0(q)= \frac12 \frac{1}{q^2 + \lambda_2 \rho_0}$ is the bare propagator.
One can see in the singular at low energy is suppressed by a factor $q^4$, i.e.,
\bea
	\Big[\chi_{\delta \rho_n \delta \rho_n}^\text{singular}\Big]''(\omega, q) =  \frac{(\omega^2-q^2)^2}{128 \rho_0 (\omega^2-q^2+ \lambda_2 \rho_0)^2} \Theta(\omega^2-q^2).
\eea
The static scalar susceptibility is given by $\chi_{\delta \rho_n \delta \rho_n}''(\omega)= \frac{\pi}{4\sqrt{\lambda_2 \rho_0}} \delta(\omega- \sqrt{\lambda_2 \rho_0})+ \frac{\omega^3}{128 \rho_0 (\omega^2-q^2+ \lambda_2 \rho_0)^2} + \text{regular terms}$, where we can see clearly that the singularity is well suppressed. As a consequence, the peak of the Higgs modes can be observed experimentally in the scalar susceptibility without covering from the IR divergence.

The suppression in above calculation is actually due to the different decomposition of the order parameter. In the amplitude-angle decomposition, $\Delta = |\Delta| e^{\theta}$, it is the two derivatives appearing in the coupling $|\Delta| (\partial \theta)^2$ that lead to a suppression. We emphasize that the Higgs mode in condensed-matter systems is not like Bogoliubov quasiparticle that can be detected by tunneling experiments. As a neutral collective mode, the Higgs mode is not easy to excite and detect. The different self-energy from the different decomposition of order parameters by perturbative calculation will not lead to a direct experimental observation. If both perturbative calculations remain qualitatively correct when high-order corrections are included, the lesson above tells us that the Higgs mode can be observed by suitable observable in experiments. For example, besides the contribution from Bogoliubov quasiparticle in a superconductor, the amplitude mode can lead to an excess contribution to the dynamical conductivity. The leading contribution from the amplitude mode is theoretically shown to have a hard gap at the frequency set by the mass of Higgs mode~\cite{arovas2011}. The experiments in disordered NbN and InO film confirm the extra contribution in dynamical conductivity near the superconductor-insulator transition~\cite{sherman2015}.

\begin{figure}
	\subfigure[]{
		\includegraphics[height=1.8cm]{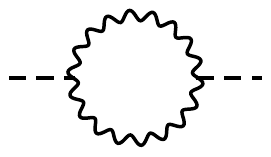}} \quad
	\subfigure[]{
		\includegraphics[height=1.8cm]{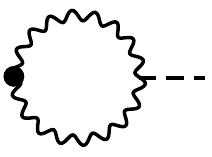}} \quad
	\subfigure[]{
		\includegraphics[height=1.8cm]{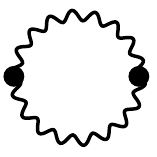}}
\caption{\label{fig2} The one-loop diagrams containing singular polarization. The dashed lines and the wavy lines represent Higgs mode propagator and Leggett mode propagator, respectively.}
\end{figure}

\end{widetext}

\end{document}